# Silicon nitride metalenses for unpolarized high-NA visible imaging

Zhi-Bin Fan[1,2,*], Zeng-Kai Shao[1,3,*], Ming-Yuan Xie[1,2], Xiao-Ning Pang[1,2], Wen-Sheng Ruan[1,2], Fu-Li Zhao[1,2], Yu-Jie Chen[1,3,†], Si-Yuan Yu[1,3,4], and Jian-Wen Dong[1,2,†]

## Abstract

As one of nanoscale planar structures, metasurface has shown excellent superiorities on manipulating light intensity, phase and/or polarization with specially designed nanoposts pattern. It allows to miniature a bulky optical lens into the chip-size metalens with wavelength-order thickness, playing an unprecedented role in visible imaging systems (e.g. ultrawide-angle lens and telephoto). However, a CMOS-compatible metalens has yet to be achieved in the visible region due to the limitation on material properties such as transmission and compatibility. Here, we experimentally demonstrate a divergent metalens based on silicon nitride platform with large numerical aperture (NA~0.98) and high transmission (~0.8) for unpolarized visible light, fabricated by a 695-nm-thick hexagonal silicon nitride array with a minimum space of 42 nm between adjacent nanoposts. Nearly diffraction-limit virtual focus spots are achieved within the visible region. Such metalens enables to shrink objects into a micro-scale size field of view as small as a single-mode fiber core. Furthermore, a macroscopic metalens with 1-cm-diameter is also realized including over half billion nanoposts, showing a potential application of wide viewing-angle functionality. Thanks to the high-transmission and CMOS-compatibility of silicon nitride, our findings may open a new door for the miniaturization of optical lenses in the fields of optical fibers, microendoscopes, smart phones, aerial cameras, beam shaping, and other integrated on-chip devices.

[1] State Key Laboratory of Optoelectronic Materials and Technologies, Sun Yat-sen University, Guangzhou 510275, China

[2] School of Physics, Sun Yat-sen University, Guangzhou 510275, China

[3] School of Electronics and Information Technology, Sun Yat-sen University, Guangzhou 510275, China

[4] Photonics Group, Merchant Venturers School of Engineering, University of Bristol, Bristol BS8 1UB, United Kingdom

*These authors contributed equally to this work.

† Correspondence: YJ Chen, E-mail: chenyj69@mail.sysu.edu.cn; JW Dong, E-mail: dongjwen@mail.sysu.edu.cn.

Over the last few years, metasurfaces have emerged a great potential to replace bulky optical components due to their ultrathin and planar features[1-6]. By arranging the array of nanoposts properly, metasurfaces can almost realize arbitrary modulation on amplitude, phase and/or polarization pixel-by-pixel in subwavelength scale, leading to plenty planar device applications including beam bending generations[7-9], holograms[10-16], wave plates[17-20], vortex beam generations[7,21-23], Bessel beam generations[24,25] and vector beam generations[26], etc. In particular, it is capable to design metasurfaces as arbitrary metalenses[24,27-47]. There are much great efforts on this fantastic high-integrated metalenses, most of which were designed at the microwave[27,28] and near infrared[24,29-37] regions. For example, a metalens with high transmission and focusing efficiency at some particular near-infrared wavelengths has been reported on Ref. 30, due to the great contributions on high-contrast dielectric nanoposts and high-precision nano-fabrication technology, instead of the early use of lossy metallic materials[24,38-42]. Very recently, it moves a crucial step that the metalenses have been designed and fabricated on low-contrast dielectric materials[22,43-47] such that their transparent window falls within visible region, inspiring many unprecedented applications with high energy availability. Among low-contrast dielectric materials, silicon nitride (SiN) films have been recognized as a fully CMOS-compatible platform for photonic devices at various wavelengths ranged from visible, telecommunication to mid-infrared region[48]. With higher linear refractive index than silicon dioxide ($SiO_2$), silicon nitride can provide strong optical confinement in small waveguide dimension, and due to its larger bandgap (~5eV) than silicon (Si), silicon nitride does not suffer from two-photon absorption and the concomitant free-carrier absorption at communications wavelengths. Furthermore, the nonlinear refractive index of silicon nitride (~2.5 × $10^{-15}$ $cm^2W^{-1}$) is about an order of magnitude larger than silica, thus a thick silicon nitride film can well confine the optical mode and is beneficial to nonlinear optics. By controlling the thickness, the waveguide dispersion can be tailored to achieve phase matching and high mode confinement. In these regards, the silicon nitride based metalens is becoming an outstanding solution with the superiority of very low optical propagation losses, full CMOS-compatibility and nonlinear manipulation.

On the other hand, numerical aperture (NA) of an optical lens system is a vital dimensionless number that characterizes the range of angles over which the system can accept or emit light, indicating the light-gathering ability as well as imaging resolution ($\lambda/2NA$). A lens with larger numerical aperture will achieve much better image performances on finer details, brighter images and shallower depth of field (DOF), etc. Currently, numerous efforts are taking place to develop the high-NA metalens[30,43-45], and some of these have already shown great image quality. For instance, the titanium dioxide metalens with NA ≈ 0.8 is used to observe some resolution charts and the high-precision resolving power has been achieved in Ref. 43. Most of the aforementioned metalenses have been designed as convergent lenses while in imaging systems, the divergent lens[41,42] also plays an important role due to its various functions on aberration correction. For instance, the conventional doublet lens is made of a concave and a convex lens to match each other, and also on the wide-angle lenses such as fish-eye and telephoto lens systems, the divergent lenses usually own much larger size and heavier weight than other convergent ones, implying a more efficient way to lighten the whole systems with the miniaturization of divergent lenses.

In this work, we design the polarization-insensitive divergent metalenses with a hexagonal silicon nitride array that consists of cylinder nanoposts in subwavelength scale. With large numerical aperture of 0.98 and high transmission of 0.8, a 100-μm-diameter micro metalens operates at 633 nm and is fabricated with 695-nm-thick silicon nitride array. Another 1-cm-diameter macroscopic metalens has also been shown, including over half billion nanoposts. After characterizing the virtual focus point resolution with nearly diffraction-limit, the micro metalens enables to miniature objects into the micro-scale size as small as the core of single-mode fiber, while the macro metalens demonstrates a great potential for wide-viewing-angle applications.

We first design a set of nanoposts with high transmittance and full 2π phase coverage in the periodic grating. At the basic nanostructural level in metalens design, there are various geometries of dielectric nanoposts including square cylinder[46], cuboid[43], slit[25,32], cylinder[44], elliptic cylinder[35], and combined meta-molecules[31,34], etc. Such nanoposts can be divided into two types by whether they are sensitive to polarization or not. The polarization-sensitive nanoposts are either spatial anisotropic or rotational-placing derived from the theory of Pancharatnam-Berry phase, while the polarization-insensitive ones are always cylinders or square cylinders, using diameters or side lengths to map the phases. For our purpose, we choose cylinder as the basic nanostructure to control the unpolarized light. To decide the phase sample method in metalens, we consider the rectangular and hexagonal lattices for simplicity. It has been pointed out that the differences in complex transmission coefficient between these two sample methods are negligible as the scattering of nanopost is a local effect and the coupling among nanoposts is quite weak[30]. However, we expatiate that for a circularly continuous bandlimited phase profile, hexagonal sampling can extend the limitation of sampling interval and require 13.4% fewer samples than rectangular sampling to represent the same phase profile, leading great merit to reduce the burden of fabrication, especially for the fabrication of macro metalenses.

Figure 1(a) illustrates the schematic of the hexagonal periodic grating constructed by a low-refractive-index silicon nitride circular nanoposts array, with a subwavelength lattice constant ($a$ = 416 nm). As the nanoposts stand on the silicon dioxide substrate, the lattice constant is less than the propagation wavelength of 633 nm in the substrate (633 nm/$n_{SiO_2}$ = 436 nm) but greater than the strict diffraction condition (633 nm/$2n_{SiO_2}$ = 218 nm). It ensures that only the zeroth order diffraction exists for the normally incident light, but the feeble first order diffraction will appear for the obliquely incident light. We use the rigorous coupled-wave analysis (RCWA)[49] to simulate the periodic gratings. In the simulations, a plane wave with y-polarization normally illuminates on the silicon nitride nanoposts from the bottom of substrate, yielding the complex transmission coefficient t and reflection coefficient r. The refractive index of silicon nitride is n ≈ 2 around the visible region. As is known that in a homogeneous film with thickness $d$, the maximum exiting phase is given by $\varphi_{max}$ = 2π (n-1) $d$ / λ, where we assume air as the ambient medium. Thus, to achieve a full 2π coverage, the thickness of silicon nitride nanoposts need to be greater than the wavelength, and here we fix it as 695 nm. By varying the diameter of nanoposts for the given lattice constant, the induced phases and transmission amplitudes are calculated. Due to the geometrical symmetry of the nanoposts, there are no differences in transmission between two orthogonal linear polarized normally incident lights, meaning the polarization-insensitivity. Figure 1(b) displays top and side views of the normalized magnetic field intensity in the periodic grating with a nanopost diameter of 200 nm. It can be seen that the magnetic field is well concentrated inside the nanoposts (white dashed lines) due to strong optical confinement of silicon nitride material, implying that each silicon nitride nanopost behaves as a weakly coupled low-quality factor resonator. Therefore, it can be regarded as an independent response, which is the fundamental assumption for metasurface design.

In order to give a complete picture to compare different dielectric materials, a parameter sweep of complex transmission coefficient is shown in figure 1(c) and 1(d) in the plane of nanopost diameter and refractive index, at a given slab thickness of 695 nm and a working wavelength of 633 nm. In such cases, it is interesting that the lower refractive index (e.g. n ~ 2, the white vertical dashes lines) can support high and smooth transmission spectrum within a relatively broad diameter range from 50 to 400 nm, except the low-transmission resonant dip when the diameter has the value of 292 nm, see the blue curve of figure 1(g) while the red curve indicates the full 2π phase coverage. In contrast, a multiple resonant area appears inside the high refractive index, i.e. the upper-right of figure 1(c), leading to poor transmission. On the other hand, Figure 1(d) illustrates the much lower refractive index not to support the full 2π phase coverage. As a result, we can choose the material of silicon nitride (n ~ 2) to achieve a full 2π phase coverage while maintaining high transmission at 633 nm. Furthermore, the spectral responses of the silicon nitride periodic gratings are also shown in figure 1(e) and 1(f). As depicted in figure 1(e), the transmission keeps up close to 1 for a wide range from 530 nm to 780 nm in visible region except a little resonant dips, while in the relatively low wavelength region, a low transmission area can be observed which will influence the usage of a metalens, see the blue curve of figure 1(h) for a fixed diameter case. This low transmission area should be avoided when choosing the discrete nanoposts in the subsequent round of design. Figure 1(f) shows that the full $2\pi$ phase coverage can be achieved in almost the whole visible region. More than $2\pi$ phase coverage in the lower wavelength area implies the much flexibility of silicon nitride for designing in lower wavelengths. Though a low transmission area also exists in the lower wavelength area as figure 1(e) shown, it is able to be removed by properly adjusting the thickness of nanoposts. In brief, we have eventually attained the silicon nitride nanopost parameters that can support high transmission in the broadband visible region and provide the full $2\pi$ phase coverage in the designed wavelength. With these parameters in hand, we then can achieve an arbitrary transmission phase profile by appropriately arranging the nanoposts in nonresonant regime.

To achieve a desired spatial phase profile, we first choose six nanoposts with different diameters to map six linear orders between 0 and $2\pi$ based on the result of figure 1(g) and on the consideration of transmission in figure 1(e). Then we discretize the continuous phase profile onto a hexagonal lattice with the subwavelength lattice constant, yielding a discrete spatial phase map. Finally, according to the discrete phase map, we place the six nanoposts appropriately on the silicon dioxide substrate in corresponding positions. The selected nanopost diameters vary from 170 nm to 374 nm while the fixed thickness is 695 nm, resulting in a maximum aspect ratio ~1/4 and a minimum space of 42 nm between adjacent nanoposts.

In our design, we choose the spatial phase profile of the metalens as a divergent lens, illustrated by figure 3(a), yielding $\varphi(x,y) = \frac{2\pi}{\lambda}\left(\sqrt{x^2 + y^2 + f^2} - |f|\right)$ where $\lambda$ = 633 nm is the designed wavelength, $f$ is the focal length, and $(x, y)$ is the in-plane coordinate of each point in the phase profile. We design the micro metalens with diameter of 100 μm and focal length of -10 μm. The NA of the metalens is as high as 0.98. Note that with the near-unity NA, the narrowest width of the outermost wave zone is near to a wavelength ($\lambda$/NA), implying that the number of quantized steps in this outermost wave zone cannot achieve to six, which may affect the performance of the metalens. In order to evaluate the performance of the designed metalens, we propose two simple and quick methods based on the fundamental assumption of independent response among nanoposts: the ideal post method and the grating cell method (see supplementary material S3). Both results conform to that in the FDTD method, and all of them in figure S2 support the only one virtual focus in the designed focal length, verifying the design feasibility of our metalens. With these previously desired evaluation results, we can carry on the next challenging fabrication process.

To achieve the full $2\pi$ phase coverage at 633 nm, silicon nitride nanoposts of 695 nm thickness are required. A diagrammatic drawing of metalens' fabrication process is shown in figure 2. In our work, the inductively coupled plasma-chemical vapor deposition (ICP-CVD) process is optimized to deposit silicon nitride film on silica substrate at a low temperature (300 ℃ or 75 ℃) with significantly decreased hydrogen concentration and controlled film stress[50,51]. After deposition, the metalens patterns are defined in a high resolution positive resist polymethylmethacrylate (a 550-nm-thick film of PMMA A7, MicroChem) by an electron-beam lithography (EBL) system (Vistec EBPG5000 ES) at 100 KV. To ensure the structure edges as smooth as possible of, we use the sequence fracturing method and the high resolution mode to approximate the nanoposts in the layout. Followed by the development of PMMA, a 100-nm-thick Cr layer is deposited by electron beam evaporation (DE400DHL, DE Technology Inc.) as the hard mask to achieve high etch selectivity. Subsequently a lift-off process combined with $O_2$ plasma cleaning is applied to strip residual patterns. Next, the reactive-ion-etch (RIE) (Oxford Instrument

Plasmalab System 100 RIE180) with a mixture of $CHF_3$ and $O_2$ gases is applied to etch through the 695-nm-thick silicon nitride layer. The RIE process has been optimized with regard to vertical and smooth sidewalls of nanoposts. Finally, the rest of Cr layer is removed by the stripping solution (Ceric ammonium nitrate) and hence silicon nitride nanoposts are patterned as the desired metalens. Scanning electron microscope (SEM) images of the 100-μm-diameter micro metalens are exhibited. Top-view SEM image of a portion of the fabricated micro metalens with a relatively high magnification are displayed in figure 3(b) and the side-view SEM image of the micro metalens in figure 3(c) describes the vertical profile of nanoposts.

In this work, we combine the rigorous vector diffraction analytical method with the scalar diffraction theory to simulate the whole effect of the micro metalens. We will have a virtual focal spot due to the negative focal length of the divergent lens. Such virtual point can be straightforward observed through additional optical microscopy (e.g. objective with proper setup), however, it cannot be acquired from full wave simulation directly, even using the 3D finite difference time domain (FDTD) method. In order to analyze the virtual focusing effect side-by-side, we employ a field-tracing method[52] on back propagating calculation to mimic the human-eye observation. We first consider a y-polarized CW Gaussian beam normally incident onto the front side of the metalens. Then the FDTD package will calculate the electromagnetic field distributions throughout the metalens, and we will have the field distribution at the back side of the metalens. Finally, the virtual focal plane and the virtual 3D field distributions can be traced back along the opposite direction of the incident light in free space, which can be done by the field-tracing method proposed in the commercial software VirtualLab Fusion.

To characterize the virtual focus spot of the divergent micro metalens, we measured the focusing effect at different wavelengths from 480 nm to 780 nm in the visible region. The measurement setup is illustrated in figure 3(d) and more details can be found in supplementary material S5. Figure 4(a) shows the intensity profile in the virtual focal planes of five different wavelengths, depicted by various types of pseudo-colors. One can see that the intensity is highly confined at the center and the spot is close to diffraction-limit within the measured wavelength region. This advantage enables for broadband imaging with high resolution. The simulation results calculated by the field-tracing method, as shown in figure 4(b), are well consistent with those in measurement, even for the feeble sidelobes around the center spots. To quantitatively evaluate the center spot, we plot out the cross sections of the spot for all the five wavelengths, and illustrate the intensity profiles in the x-direction at 480, 633, 780 nm in figure 4(c-e). The remaining results are collected in figure S5. The dots represent the measurement while the blue and red curves denote the Airy fit using the *jinc* function and the simulation results. A prominent feature is to gain a nearly diffraction-limited virtual focus spot at 633 nm, by verifying the value of the full width at half-maximum (FWHM) to be 364 nm, very close to the simulation value of 357 nm. The nearly diffraction-limited spot are also found in other wavelengths of 480, 532, 730, and 780 nm, with the differences within 4.5% between measurement and simulation. It is noted that the FWHM values do not increase monotonically from blue to red region and it reaches the minimum at the designed wavelength of 633 nm, deviating from that of traditional diffraction optical micro-elements and leading to some unexpected chromatic aberration in optical imaging.

Figure 5 depicts the focus spots of the divergent micro metalens along the propagation z-direction at five different wavelengths. The focus spots are observed within the negative z position while the micro metalens is placed on the plane of z = 0. This is clearly seen from the consistent results between the measurement in figure 5(a) and the simulation in figure 5(b), verifying the diverging feature of the metalens. At the center wavelength of 633 nm, e.g. the third row in figure 5(a) and 5(b), there is a remarkable feature such that only one focus spot can be observed at the distance of -10 μm, when inserting the divergent micro metalens with its diameter of 100 μm. This is a strong evidence to claim the realization of nearly-unity NA metalens in air, with the value as high as 0.98. The measured focus spot has little position deviation from the simulation, as shown in figure 5(c), but this will not affect the lateral imaging quality. For the other wavelengths beyond 633 nm, on the other hand, several secondary focal points exist along the propagation direction, locating in the negative z axis and at different distances from the metalens. In particular, such secondary focal points of shorter wavelength are further away from the metalens than those of longer ones. It leads to a slender tail around the main focal point, bringing additional monochromatic aberrations for imaging and negative influence on the device. Furthermore, figure 5(d) illustrates that the focal length decreases from blue to red regions, with the same tendency as the conventional diffraction dispersion of binary optics. But such chromatic dispersion in the metalens is a little more serious compared to the conventional binary optics with an expression of $f = \lambda_{center} f_{center} / \lambda$, where the subscript stands for the center wavelength. This can be found in figure 5(d) that both measurement (blue) and simulation (red) curves have larger slope than the conventional case (dashed). Since the conventional dispersion relation is deduced from the scalar diffraction theory, this deviation also verifies that the scalar diffraction theory cannot work well in the subwavelength scale such that we must use a rigorous vector diffraction analytical method (e.g. FDTD) to simulate the inner field of metalens. Note that there is also slight difference between measurement and simulation except for the center wavelength. The chromatic dispersion in the metalens is mainly due to the different dispersion response of nanoposts and the coupling between adjacent nanoposts at various wavelengths, which may also cause extra aberrations. The transmission spectrum is also measured by a homemade transmission setup (figure S3), and it comes up to ~80% high within several visible regions, manifesting the advantage of high transparency through the silicon nitride material.

In order to demonstrate the ability for practical imaging, we employ the above divergent metalens to image a logo "SYSU" with the actual physical size of $300 \times 340$ μm$^2$. The measurement configuration schematic is drawn in figure 6(a). A white light source (Thorlabs OSL2) is applied to illuminate the logo "SYSU" through a pin hole and a series of narrow-band (10 nm) filters (not shown in the light path) with different center wavelengths are placed between the lamp and the pinhole. An attenuator is used to control the intensity of the incident light. To conveniently adjust the relative distance between the object and the metalens, we first use a 40× objective (GCO-21) to gather the rays from the logo, so as to obtain the first size-reduced logo image as the object of the metalens. Such object is then secondly imaged by the divergent metalens to form a much reduced virtual image, which is finally observed by an imaging microscopy system consisting of a 100× objective (Olympus MPLFLN100xBD), a tube lens (Thorlabs ITL200) and a colorful CCD camera (Mshot MC20). Figure 6(b) plots out the schematic diagram in a much simple format. Note that the logo and the 40× objective are mounted on the same multi-axis translation stages so that once the distance between the logo and the 40× objective is determined, they will be moved together to ensure that the object will not change except for its position.

One can see in figure 6(c), that the object with the size of ~23 ×28 μm$^2$ is the reduced image of the original logo "SYSU" through the 40× objective at 635 nm. After inserting the divergent metalens, a more reduced virtual image is captured by the microscopy system. In the measurement, such virtual image finally has a size of ~3 × 3.6 μm$^2$, shrinking about 1/59 of the object size. The whole object has been reduced into the ultra-small circular field-of-view highlighted by the red dash circle in figure 6(d), implying a potential application of wide field-of-view imaging. However, the strong shrinking ability will somehow degrade imaging resolution due to monochromatic aberrations. For example, the letters "SYSU" boundaries become blurred after metalens reduction, compared with the much sharper boundary without metalens. Such effects may come from several aspects. First is the sample rate which is not high enough as the numerical aperture of the metalens reaches to nearly-unity. The undersampling gives rise to that the discrete spatial phase profile cannot be provided to reconstruct the continuous spatial phase profile perfectly, causing exact aberrations. Second is that the responses of all selected nanoposts in the metalens are not completely consistent with that in the periodic grating, leading to the distortion of spatial frequency spectrum. Moreover, the defects of fabrication and measurement error may also increase the aberrations.

In other word, one may obtain an ultra-wide field-of-view by the high NA divergent metalens, but it will lose some details of the object. The way to avoid losing details may be to reduce the NA of the metalens so to increase the required sampling interval, or to break through the Rayleigh diffraction limit to provide much more resolving power. The more entertaining approaches are to optimize the sample method using the modern optimization theories and to further challenge the fabrication limits. Figure 6(e, f) and 6(g, h) show the corresponding images at 485 nm and 542 nm, respectively. These results are similar to those at 635 nm. All letters in both images can still be clearly recognized, despite the strong reduction by the metalens, showing the high resolution at these wavelengths. Note that the final image can be reduced to a size of about $3 \times 3.6$ μm$^2$, very close to the size of a piece of single-mode optical fiber. It means that if the metalens would mount on the fiber end face, one might use such meta-fiber to image objects with wide field-of-view so that more information can be gained than before. Furthermore, as the spatial phase profile of the metalens can be designed arbitrarily, it may inspire an interesting idea to design a metalens with multiple focal points[37] in different position but the same focal plane. It can be used for the optical fiber bundle and provide a more excellent performance for endoscope imaging, biosensor and biomedical research.

Based on the above considerations, we further fabricate a macro metalens with diameter of 1 cm and focal length of -4 mm, yielding a little lower NA≈0.78. Figure 7(a) shows the photograph of the actual sample in which the ruler is used to emphasize the 1-cm-diameter metalens. To display its wide field-of-view application, we first use the macro metalens to watch a logo directly using the measurement setup with the 10× objective in figure S4. Figure 7(b) and (c) respectively show the original objects without the macro metalens and the corresponding images with the macro metalens. It is clear that in the same circular area in the CCD camera, one can get much more field-of-view in figure 7(c) than that in figure 7(b), indicating a powerful evidence that the macro metalens broadens the system's horizon. Note that the number of nanoposts in the 1-cm-diameter macroscopic metalens is over half billion, which means that it need to consume large resources (including preparation time and equipment) to fabricate. To optimize the sample method so as to reduce the sampling number while keeping the high imaging performance is very important. The feasible plans may include applying the deep learning method[53] and some high speed analytical algorithms with the Graphics Processing Unit (GPU) platform just as holographic computations done before[54].

In conclusions, we have demonstrated the low-contrast dielectric metasurface as a divergent metalens with near-unity NA and high transmission for unpolarized visible light by using the hexagonal array of cylindrical nanoposts based on the CMOS-compatible silicon nitride platform. On the aspect of fabrication, we have firstly produced a near-unity NA micro metalens with 100-μm-diameter and 695-nm-thick silicon nitride nanopost array, which have a maximum aspect ratio ~1/4 and a minimum space of 42 nm between adjacent nanoposts. The experiment results are well consistent with the early designs, showing the unique virtual focal spot with nearly diffraction-limited FWHM to declare the effective near-unity NA and high resolving power of the micro metalens. Furthermore, a macroscopic high-NA metalens with 1-cm-diameter has also been fabricated, which consists of over half billion nanopost unit cells. The imaging results of each metalens have well demonstrated the wide field-of-view potential. Consequently, combined with the chip-size feature and wavelength-scale thickness, the micro metalens is able to be used for integration in microscopic optical systems like optical fibers for biophotonics. As the divergent metalens can provide wide field-of-view for imaging, it may allow the miniaturization of endoscopes. Moreover, it can be applied to some macroscopic optical devices like smart phones and telescopes by fabricating the much larger metalens. Most importantly, with the CMOS-compatibility of the silicon nitride material and its superiority in nonlinear optics (e.g. broad-bandwidth phase matching), it may inspire us to integrate metalenses into CMOS devices and pave the way for other on-chip optical interconnections. We envision more applications of high-NA silicon nitride metalenses in realization of more complex optical systems.

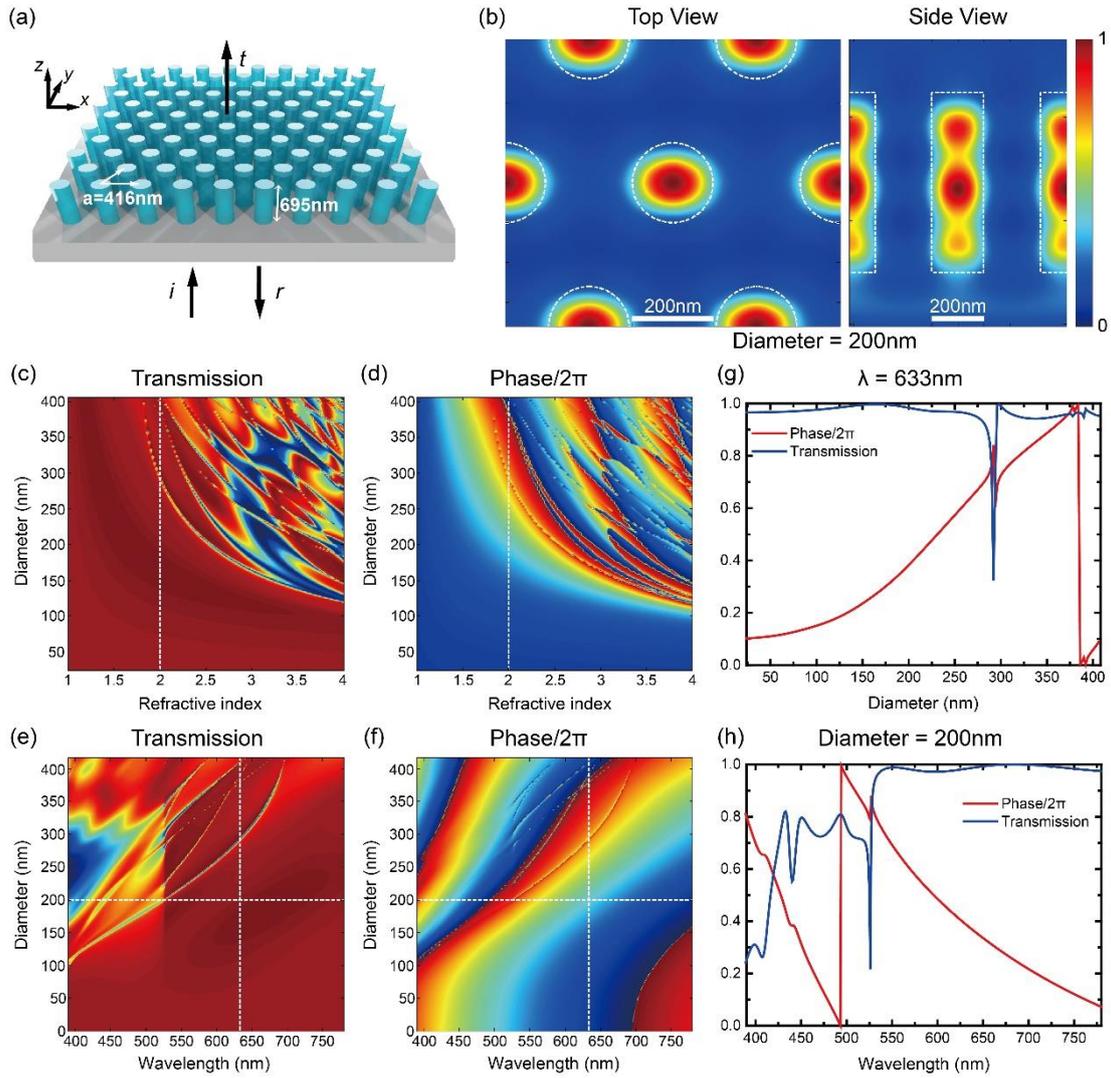

**Figure 1**. Simulation results of silicon nitride periodic gratings. (**a**) Schematic of the periodic grating constructed by a low-refractive-index silicon nitride array with circular nanoposts on a fused silica substrate. The hexagonal lattice constant is a = 416 nm, and the thickness of nanoposts is 695 nm. (**b**) Top and side views of the normalized magnetic field intensity with the nanopost diameter of 200 nm. The white dashed frame depicts the boundaries of the silicon nitride nanoposts. A plane wave with y-polarization is normally incident onto the silicon nitride nanoposts from the bottom of substrate. Scale bar, 200 nm. (**c-d**) Calculated transmission and its phase as a function of the refractive index and the post diameter at λ = 633 nm. (**e-f**) Calculated transmission and its phase as a function of the wavelength and the post diameter with refractive index n = 2. (**g**) The profile of the vertical dashed lines in (**c-f**), showing the full 2π phase coverage and the close-to-1 transmission spectrum for a family of silicon nitride periodic gratings at λ = 633 nm. (**h**) The profile of the horizontal dashed lines in (**e-f**), illustrating the spectral response for silicon nitride periodic gratings with the nanopost diameter of 200 nm and the high efficiency in the broadband visible region.

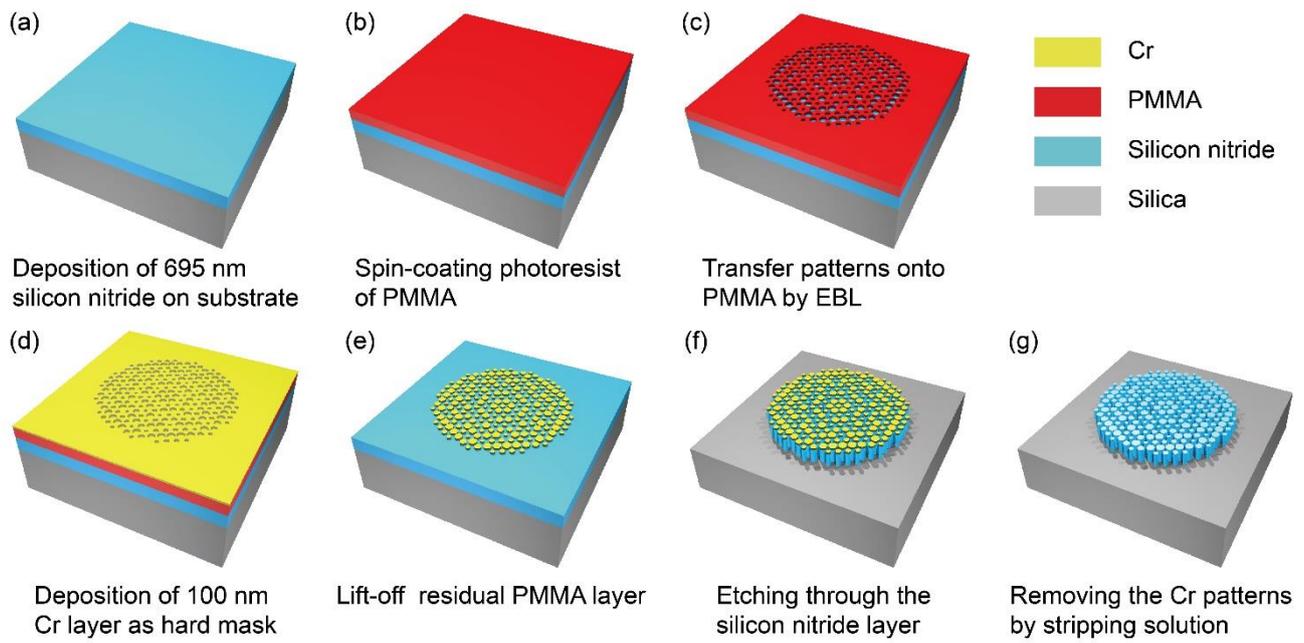

**Figure 2**. Schematic illustration of metalenses fabrication. (a) Deposition of 695-nm-thick silicon nitride on silica substrate. (b) Spin-coating photoresist of PMMA. (c) Transfer patterns onto PMMA by EBL. (d) Deposition of 100-nm-thick Cr layer as hard mask. (e) Lift-off the residual PMMA layer. (f) Etching through the 695-nm-thick silicon nitride layer. (g) Removing the Cr patterns by stripping solution.

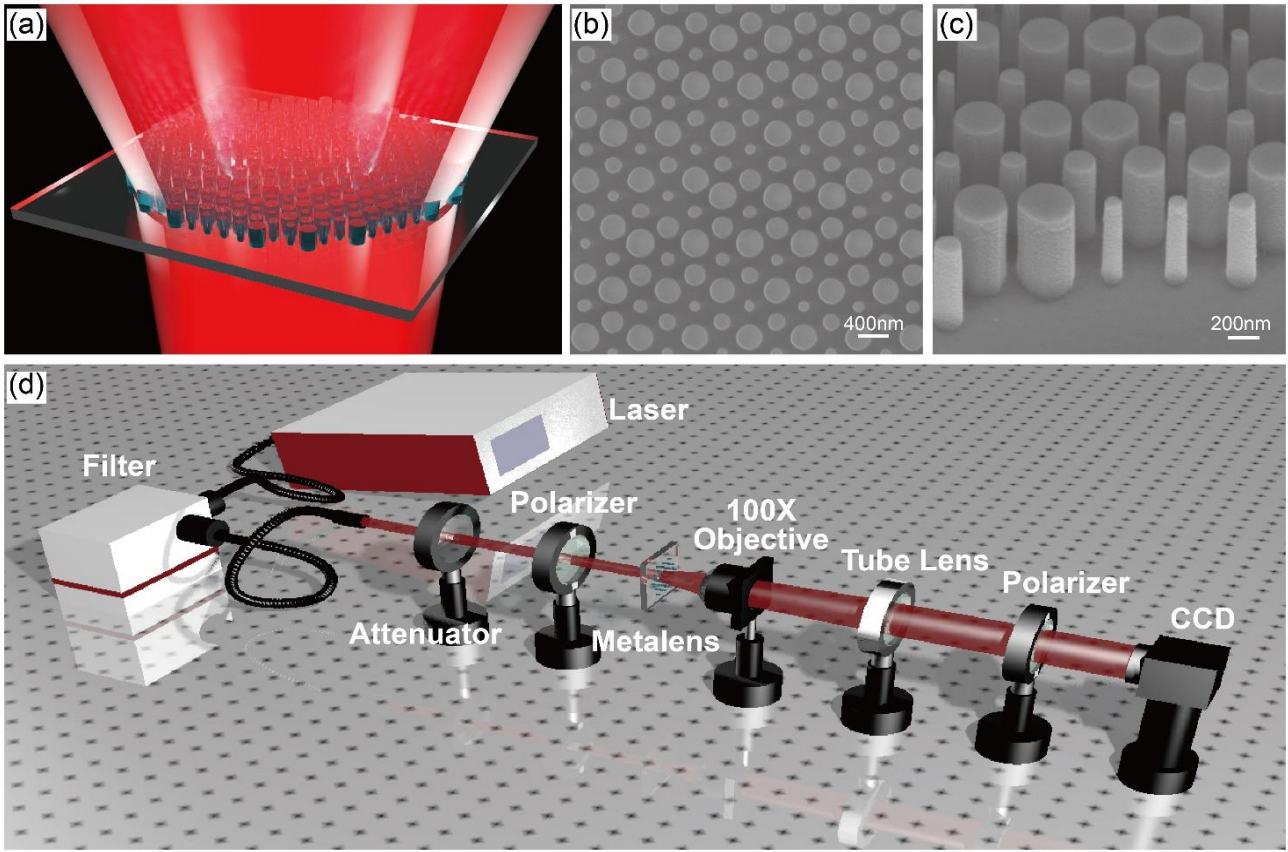

**Figure 3**. Schematic diagram, SEM images and measurement setup of the 100-μm-diameter micro metalens. (**a**) Schematic diagram of a divergent metalens operating in the transmission mode. (**b**) Top-view SEM image of a portion of the fabricated micro metalens. Scale bar, 400 nm. (**c**) Side-view SEM image of a portion of the fabricated micro metalens with a relatively high magnification. Scale bar, 200 nm. (**d**) Measurement setup for characterizing the chromatic dispersion features of the micro metalens at different wavelengths.

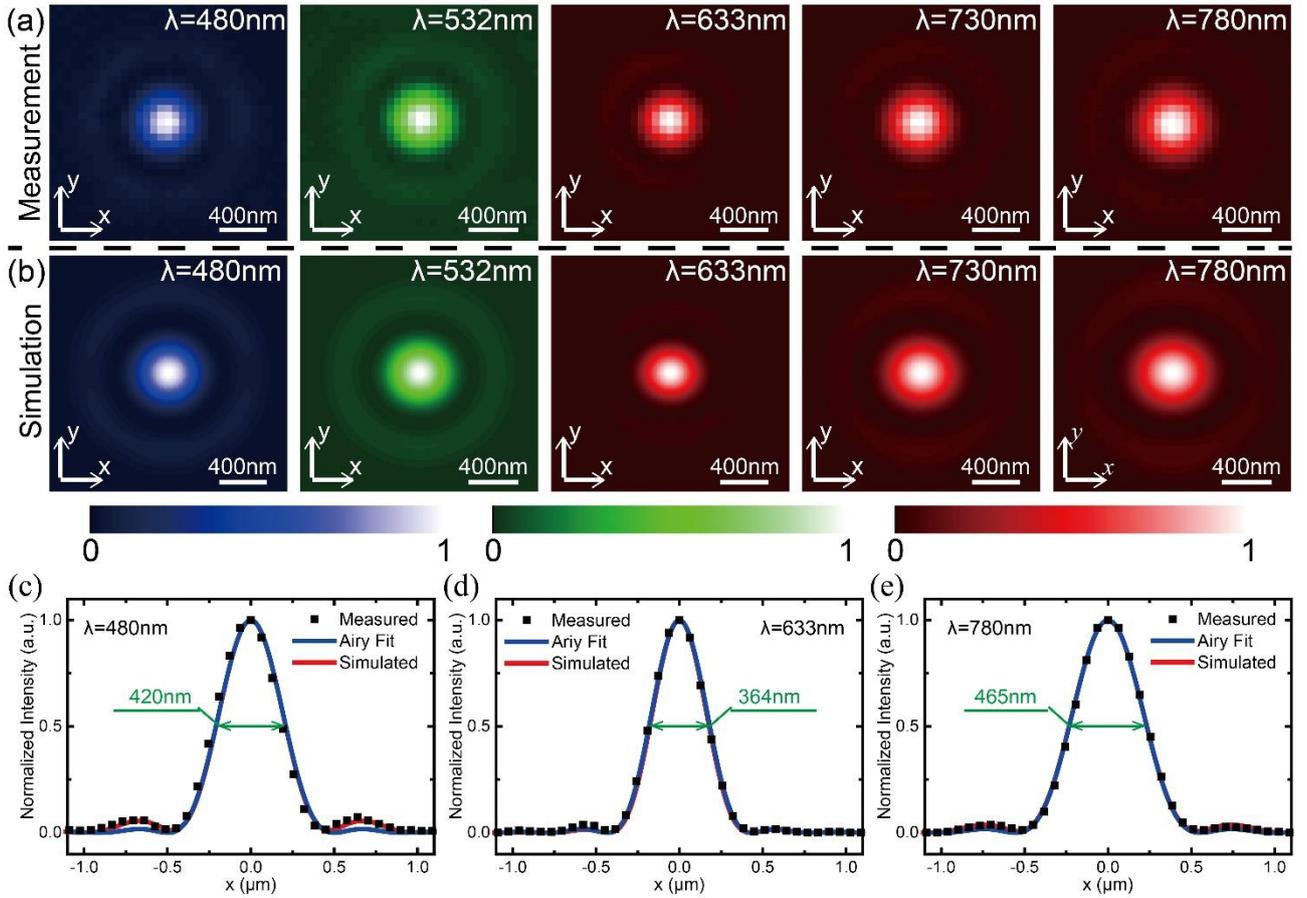

**Figure 4.** The virtual focal planes of the micro metalens. (**a-b**) Measurement and simulation of intensity patterns in the virtual focal planes of the micro metalens at five different wavelengths of 480, 532, 633, 730, and 780 nm. Three kinds of pseudo colors are used for color bars. Scale bar, 400 nm. (**c-e**) The x-direction cross sections of the measured and simulated intensity profiles at three wavelengths. The red solid line denotes the simulated results while the blue solid line is the Airy fit of measured data (black square point) using the *jinc* function. The green texts give the full width at half-maximums (FWHMs) of fitting measured data, exposing the nearly diffraction-limited virtual focus spots.

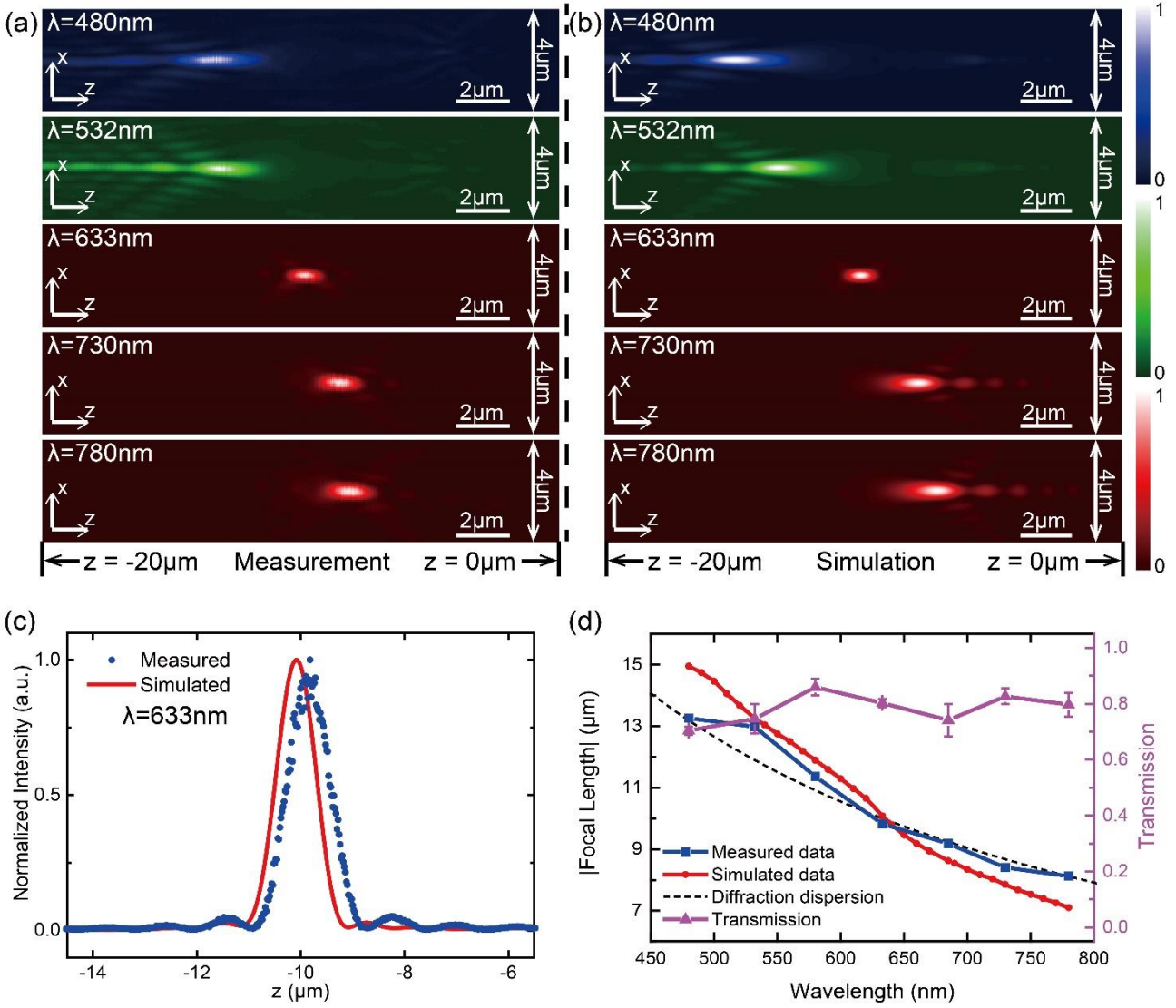

**Figure 5**. The x-z planes of the micro metalens in different wavelengths. (**a-b**) Normalized measured and simulated intensity distributions in x-z planes at different wavelengths respectively. Three kinds of pseudo colors are used for color bars. Scale bar, 2 μm. (**c**) The z-direction cross sections at 633 nm. The red solid line is the simulated data and the blue circle points are measured data, showing no significant difference with simulated data except for little position deviation of focus. (**d**) Measured and simulated virtual focal lengths of the micro metalens as a function of wavelength. The black dash line represents the diffraction dispersion curve which follows with an expression of $f = \lambda_{center} f_{center} / \lambda$, where the subscript stands for the center wavelength. It notes that the chromatic dispersion of the metalens is more serious than diffraction dispersion. The pink curve demonstrates the measured transmission spectrum, which comes up to ~80% high within several visible regions. The error bars represent the s.d. for three measurement repetitions.

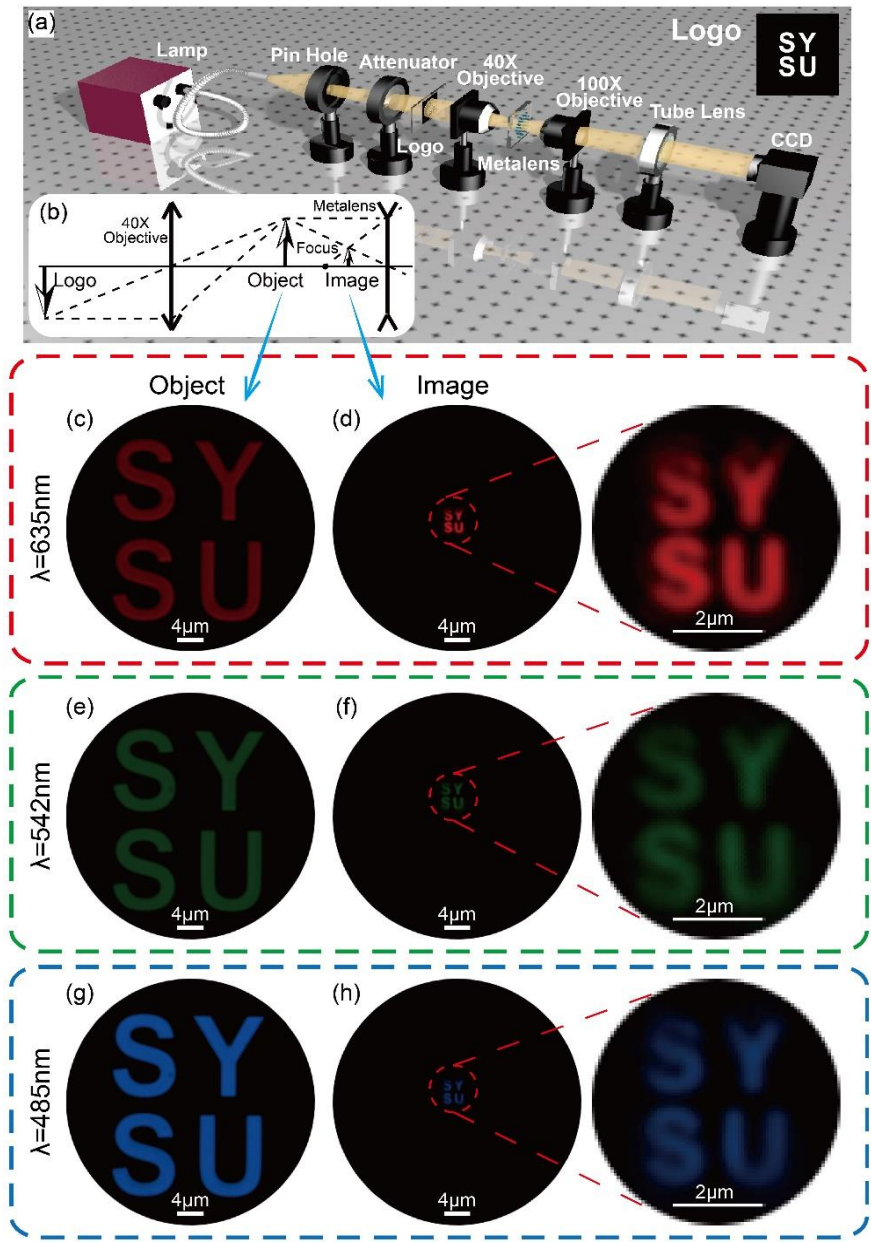

**Figure 6**. Image demonstrations of the micro metalens. (**a**) Measurement configuration for image demonstrations of the micro metalens. (**b**) Simple schematic diagram of imaging mechanism among the logo, the 40×objective and the micro metalens. (**c**) The object illuminated by inserting narrow-band (10 nm) filters of 635 nm between the lamp and pinhole. Scale bar, 4 μm. (**d**) The corresponding image of (**c**) observed by the micro metalens and its zoom-in view of the red dashed circular area. The whole object has been reduced to the small circular field-of-view, implying a wide field-of-view potential for imaging. (**e-h**) The corresponding results at wavelength of 485 nm and 542 nm.

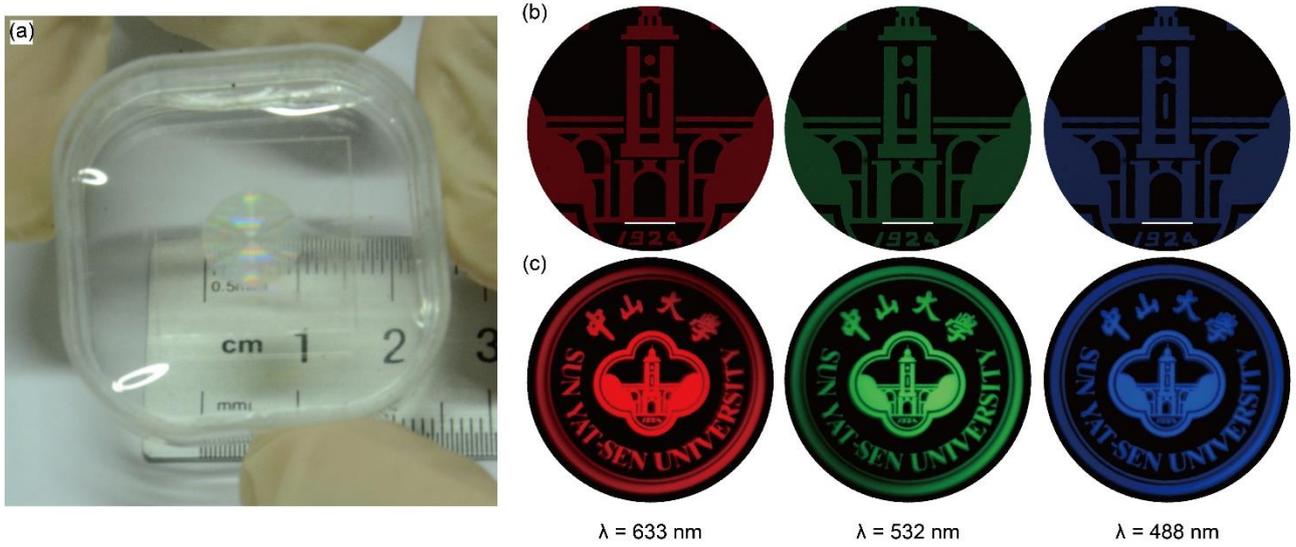

**Figure 7**. Image demonstrations of the macro metalens. (**a**) Photograph of the macro metalens with focal length of -4 mm, in which the ruler is used to emphasize its 1-cm-diameter size. The number of nanoposts in such macroscopic metalens is over half billion due to the subwavelength lattice constant (416 nm). (**b**) The central parts of object logos observed by the 10× objective without the macro metalens at three different wavelengths, respectively. Scale bar, 100 μm. (**c**) The corresponding images extracted in the same areas of (**b**) on the CCD camera, directly showing the wider field-of-view by achieving the whole logo.

# REFERENCES

1  Kildishev, A. V., Boltasseva, A. & Shalaev, V. M. Planar Photonics with Metasurfaces. *Science* 2013; **339**.
2  Yu, N. & Capasso, F. Flat optics with designer metasurfaces. *Nat Mater* 2014; **13**: 139-150.
3  Shaltout, A. M., Kildishev, A. V. & Shalaev, V. M. Evolution of photonic metasurfaces: from static to dynamic. *J. Opt. Soc. Am. B* 2016; **33**: 501-510.
4  Hou-Tong, C., Antoinette, J. T. & Nanfang, Y. A review of metasurfaces: physics and applications. *Reports on Progress in Physics* 2016; **79**: 076401.
5  Jahani, S. & Jacob, Z. All-dielectric metamaterials. *Nat Nano* 2016; **11**: 23-36.
6  Zhang, L., Mei, S., Huang, K. & Qiu, C.-W. Advances in Full Control of Electromagnetic Waves with Metasurfaces. *Advanced Optical Materials* 2016; **4**: 818-833.
7  Yu, N., Genevet, P., Kats, M. A., Aieta, F., Tetienne, J.-P. *et al.* Light Propagation with Phase Discontinuities: Generalized Laws of Reflection and Refraction. *Science* 2011; **334**: 333.
8  Ni, X., Emani, N. K., Kildishev, A. V., Boltasseva, A. & Shalaev, V. M. Broadband Light Bending with Plasmonic Nanoantennas. *Science* 2012; **335**: 427.
9  Zhou, Z., Li, J., Su, R., Yao, B., Fang, H. *et al.* Efficient Silicon Metasurfaces for Visible Light. *ACS Photonics* 2017; **4**: 544-551.
10 Chen, W. T., Yang, K.-Y., Wang, C.-M., Huang, Y.-W., Sun, G. *et al.* High-Efficiency Broadband Meta-Hologram with Polarization-Controlled Dual Images. *Nano Letters* 2014; **14**: 225-230.
11 Ni, X., Kildishev, A. V. & Shalaev, V. M. Metasurface holograms for visible light. 2013; **4**: 2807.
12 Wen, D., Yue, F., Li, G., Zheng, G., Chan, K. *et al.* Helicity multiplexed broadband metasurface holograms. 2015; **6**: 8241.
13 Zheng, G., Mühlenbernd, H., Kenney, M., Li, G., Zentgraf, T. *et al.* Metasurface holograms reaching 80% efficiency. *Nat Nano* 2015; **10**: 308-312.
14 Wang, B., Dong, F., Li, Q.-T., Yang, D., Sun, C. *et al.* Visible-Frequency Dielectric Metasurfaces for Multiwavelength Achromatic and Highly Dispersive Holograms. *Nano Letters* 2016; **16**: 5235-5240.
15 Li, L., Jun Cui, T., Ji, W., Liu, S., Ding, J. *et al.* Electromagnetic reprogrammable coding-metasurface holograms. *Nature Communications* 2017; **8**: 197.
16 Walter, F., Li, G., Meier, C., Zhang, S. & Zentgraf, T. Ultrathin Nonlinear Metasurface for Optical Image Encoding. *Nano Letters* 2017; **17**: 3171-3175.
17 Zhao, Y. & Alù, A. Manipulating light polarization with ultrathin plasmonic metasurfaces. *Physical Review B* 2011; **84**: 205428.
18 Yu, N., Aieta, F., Genevet, P., Kats, M. A., Gaburro, Z. *et al.* A Broadband, Background-Free Quarter-Wave Plate Based on Plasmonic Metasurfaces. *Nano Letters* 2012; **12**: 6328-6333.
19 Zhao, Y. & Alù, A. Tailoring the Dispersion of Plasmonic Nanorods To Realize Broadband Optical Meta-Waveplates. *Nano Letters* 2013; **13**: 1086-1091.
20 Ding, F., Wang, Z., He, S., Shalaev, V. M. & Kildishev, A. V. Broadband High-Efficiency Half-Wave Plate: A Supercell-Based Plasmonic Metasurface Approach. *ACS Nano* 2015; **9**: 4111-4119.
21 Yang, Y., Wang, W., Moitra, P., Kravchenko, I. I., Briggs, D. P. *et al.* Dielectric Meta-Reflectarray for Broadband Linear Polarization Conversion and Optical Vortex Generation. *Nano Letters* 2014; **14**: 1394-1399.
22 Zhan, A., Colburn, S., Trivedi, R., Fryett, T. K., Dodson, C. M. *et al.* Low-Contrast Dielectric Metasurface Optics. *ACS Photonics* 2016; **3**: 209-214.
23 Mehmood, M. Q., Mei, S., Hussain, S., Huang, K., Siew, S. Y. *et al.* Visible-Frequency Metasurface for Structuring and Spatially Multiplexing Optical Vortices. *Advanced Materials* 2016; **28**: 2533-2539.
24 Aieta, F., Genevet, P., Kats, M. A., Yu, N., Blanchard, R. *et al.* Aberration-Free Ultrathin Flat Lenses and Axicons at Telecom Wavelengths Based on Plasmonic Metasurfaces. *Nano Letters* 2012; **12**: 4932-4936.
25 Lin, D., Fan, P., Hasman, E. & Brongersma, M. L. Dielectric gradient metasurface optical elements. *Science* 2014; **345**: 298.
26 Chen, H., Chen, Z., Li, Q., Lv, H., Yu, Q. *et al.* Generation of vector beams based on dielectric metasurfaces. *Journal of Modern Optics* 2015; **62**: 638-643.
27 Xu, H.-X., Ma, S., Luo, W., Cai, T., Sun, S. *et al.* Aberration-free and functionality-switchable meta-lenses based on tunable metasurfaces. *Applied Physics Letters* 2016; **109**: 193506.
28 Chen, K., Feng, Y., Monticone, F., Zhao, J., Zhu, B. *et al.* A Reconfigurable Active Huygens' Metalens. *Advanced Materials* 2017; **29**: 1606422.
29 West, P. R., Stewart, J. L., Kildishev, A. V., Shalaev, V. M., Shkunov, V. V. *et al.* All-dielectric subwavelength metasurface focusing lens. *Opt. Express* 2014; **22**: 26212-26221.
30 Arbabi, A., Horie, Y., Ball, A. J., Bagheri, M. & Faraon, A. Subwavelength-thick lenses with high numerical apertures and large efficiency based on high-contrast transmitarrays. *Nature Communications* 2015; **6**: 7069.

31  Arbabi, E., Arbabi, A., Kamali, S. M., Horie, Y. & Faraon, A. Multiwavelength polarization-insensitive lenses based on dielectric metasurfaces with meta-molecules. *Optica* 2016; **3**: 628-633.
32  Khorasaninejad, M., Aieta, F., Kanhaiya, P., Kats, M. A., Genevet, P. *et al.* Achromatic Metasurface Lens at Telecommunication Wavelengths. *Nano Letters* 2015; **15**: 5358-5362.
33  Aieta, F., Kats, M. A., Genevet, P. & Capasso, F. Multiwavelength achromatic metasurfaces by dispersive phase compensation. *Science* 2015; **347**: 1342.
34  Shrestha, S., Overvig, A. & Yu, N. in *Conference on Lasers and Electro-Optics.* FM1H.3 (Optical Society of America).
35  Arbabi, E., Arbabi, A., Kamali, S. M., Horie, Y. & Faraon, A. High efficiency double-wavelength dielectric metasurface lenses with dichroic birefringent meta-atoms. *Opt. Express* 2016; **24**: 18468-18477.
36  Arbabi, E., Arbabi, A., Kamali, S. M., Horie, Y. & Faraon, A. Controlling the sign of chromatic dispersion in diffractive optics with dielectric metasurfaces. *Optica* 2017; **4**: 625-632.
37  Rongzhen, L., Fei, S., Yongxuan, S., Wei, W., Lie, Z. *et al.* Broadband, high-efficiency, arbitrary focusing lens by a holographic dielectric meta-reflectarray. *Journal of Physics D: Applied Physics* 2016; **49**: 145101.
38  Verslegers, L., Catrysse, P. B., Yu, Z., White, J. S., Barnard, E. S. *et al.* Planar Lenses Based on Nanoscale Slit Arrays in a Metallic Film. *Nano Letters* 2009; **9**: 235-238.
39  Chen, X., Chen, M., Mehmood, M. Q., Wen, D., Yue, F. *et al.* Longitudinal Multifoci Metalens for Circularly Polarized Light. *Advanced Optical Materials* 2015; **3**: 1201-1206.
40  Ni, X., Ishii, S., Kildishev, A. V. & Shalaev, V. M. Ultra-thin, planar, Babinet-inverted plasmonic metalenses. *Light Sci Appl* 2013; **2**: e72.
41  Chen, X., Huang, L., Mühlenbernd, H., Li, G., Bai, B. *et al.* Dual-polarity plasmonic metalens for visible light. *Nature Communications* 2012; **3**: 1198.
42  Chen, X., Huang, L., Mühlenbernd, H., Li, G., Bai, B. *et al.* Reversible Three-Dimensional Focusing of Visible Light with Ultrathin Plasmonic Flat Lens. *Advanced Optical Materials* 2013; **1**: 517-521.
43  Khorasaninejad, M., Chen, W. T., Devlin, R. C., Oh, J., Zhu, A. Y. *et al.* Metalenses at visible wavelengths: Diffraction-limited focusing and subwavelength resolution imaging. *Science* 2016; **352**: 1190.
44  Khorasaninejad, M., Zhu, A. Y., Roques-Carmes, C., Chen, W. T., Oh, J. *et al.* Polarization-Insensitive Metalenses at Visible Wavelengths. *Nano Letters* 2016; **16**: 7229-7234.
45  Chen, W. T., Zhu, A. Y., Khorasaninejad, M., Shi, Z., Sanjeev, V. *et al.* Immersion Meta-Lenses at Visible Wavelengths for Nanoscale Imaging. *Nano Letters* 2017; **17**: 3188-3194.
46  Khorasaninejad, M., Shi, Z., Zhu, A. Y., Chen, W. T., Sanjeev, V. *et al.* Achromatic Metalens over 60 nm Bandwidth in the Visible and Metalens with Reverse Chromatic Dispersion. *Nano Letters* 2017; **17**: 1819-1824.
47  Groever, B., Chen, W. T. & Capasso, F. Meta-Lens Doublet in the Visible Region. *Nano Letters* 2017; **17**: 4902-4907.
48  Moss, D. J., Morandotti, R., Gaeta, A. L. & Lipson, M. New CMOS-compatible platforms based on silicon nitride and Hydex for nonlinear optics. *Nat Photon* 2013; **7**: 597-607.
49  Liu, V. & Fan, S. S4 : A free electromagnetic solver for layered periodic structures. *Computer Physics Communications* 2012; **183**: 2233-2244.
50  Shao, Z., Chen, Y., Chen, H., Zhang, Y., Zhang, F. *et al.* Ultra-low temperature silicon nitride photonic integration platform. *Opt. Express* 2016; **24**: 1865-1872.
51  Yin, C., Chen, Y., Jiang, X., Zhang, Y., Shao, Z. *et al.* Realizing topological edge states in a silicon nitride microring-based photonic integrated circuit. *Opt. Lett.* 2016; **41**: 4791-4794.
52  Wyrowski, F. & Kuhn, M. Introduction to field tracing. *Journal of Modern Optics* 2011; **58**: 449-466.
53  Malkiel, I., Nagler, A., Mrejen, M., Arieli, U. & Wolf, L. S., Haim. Deep Learning for Design and Retrieval of Nano-photonic Structures. *arxiv1702.07949* 2017.
54  Liu, Y.-Z., Dong, J.-W., Pu, Y.-Y., Chen, B.-C., He, H.-X. *et al.* High-speed full analytical holographic computations for true-life scenes. *Opt. Express* 2010; **18**: 3345-3351.